\documentclass[12pt]{JHEP3}
\usepackage{amsfonts}

\usepackage{amsmath}

\title{Non-local charges on $AdS_5 \times S^5$ and PP-waves}
\preprint{\hepth{0310146}\\SISSA 91/2003/EP}
\author{L. F. Alday
\\
International School for Advanced Studies and INFN, Trieste, Italy. \\
The Abdus Salam ICTP, Trieste, Italy.\\
Email:alday@he.sissa.it}

\abstract{We show the existence of an infinite set of non-local
classically conserved charges on the Green-Schwarz closed
superstring in a pp-wave background. We find that these charges
agree with the Penrose limit of non-local classically conserved
charges recently found for the $AdS_5 \times S^5$ Green-Schwarz
superstring. The charges constructed in this paper could help to
understand the role played by these on the full  $AdS_5 \times
S^5$ background.}

\begin{document}


\section{Introduction}


The AdS/CFT correspondence \cite{malda1} expresses the equivalence
between type IIB string theory compactified in the background
$AdS_5 \times S^5$ and pure ${\cal N}=4$ SYM gauge theory in four
dimensions. In order to gain a better understanding of such
duality, it is of interest to exactly solve either side of the
correspondence.

In the last year, clues for the existence of integrable structures
on both sides of the duality have been pointed out. For instance,
hints of the integrability of ${\cal N}=4$ SYM in the large $N$
limit were given by studying the dilatation operator in
perturbation theory \footnote{In fact the integrability of one-loop multi-particle renormalization was discovered first in QCD
five years ago in a serie of papers \cite{int}} and it was determined that
the one loop mixing matrix for anomalous dimensions can be
identified with the Hamiltonian of an integrable spin chain \cite{dilatation}.

On the other hand, from the string theory side, after some clue
from the bosonic theory \cite{msw1} , Bena {\it et. al.} found an
infinite set of non-local classically conserved charges for the
Green-Schwarz superstring on  $AdS_5 \times S^5$
\cite{Bena:2003wd}. \footnote{This kind of charges exists also
for the pure spinor superstring \cite{Berkovits:2000fe}
\cite{val}.} This would imply that the world-sheet theory is a
integrable system.

To understand the structure of such charges, their gauge theory
dual, etc, and, eventually, to use them in order to solve string
theory on  $AdS_5 \times S^5$ is perhaps a very difficult task. A
warming up exercise would be to study the much simpler problem of
such charges on the pp-wave background. This background is simple
enough to be an exact solution of string theory, so one should be
able to understand these charges, their dual, how to use them to
solve the theory, etc, in this limit. On the other hand, the
pp-wave background is rich enough to admit infinite number of
classical conserved non-local charges and, apparently, to provide
us with some non trivial information about the role of these in
the $AdS_5 \times S^5$ background.

The aim of this paper is to study the structure of these charges
and give an algorithm that allows us to write the explicit form of
these at arbitrary order. We do it in two different ways: first,
by constructing a set of non-local charges from the world-sheet
sigma model of closed string theory on pp-waves backgrounds \footnote{Somehow related, the symmetries of D-branes on pp-waves backgrounds was studied in \cite{st} where it was determined that the open string admits an infinite number of non-local symmetries.}. Since
we need to impose periodic boundary conditions we have to take the
trace (or some invariant cyclic operator) of the product of
generators, of the Lie algebra, appearing in the charge. As the
algebra of the pp-wave is non semi simple, by just taking the
trace as the invariant we obtain trivial charges, since the trace
is degenerate. For this algebra we have, however, a non degenerate
form, that can be used in order to obtain non trivial conserved
charges. The second method consists of constructing the charges on
$AdS_5 \times S^5$ and then taking their Penrose limit. In this
case we can take the trace as the invariant, since the algebra is
semi-simple. We show, up to the first non trivial order, that both
results agree.

When considering an uncompactified sigma model the first non-local
charges generate (under repeated Poisson-Dirac brackets) an
infinite dimensional algebra, called the Yangian, and so we
generate in this way the complete tower of non-local charges
\cite{curyan}. Such algebra appears also when considering
operators acting on spin chains. A relation between both
approaches was found in \cite{Dolan:2003uh}. When one considers
periodic boundary conditions the structure of the charges changes,
and it is not clear whether one can generate the complete tower
from a finite number of charges. Indeed, for the first order
charges this is not possible.

This paper is organized as follows, in the next section we give a
brief introduction to the construction of non-local classical
conserved charges for the Green-Schwarz superstring on $AdS_5
\times S^5$ and argue that such charges should also exist for the
pp-wave. In section 3 we show how write the explicit form of these
charges in the light cone gauge and we do it for the first non
trivial orders. In section 4 we write the explicit form for the
first non trivial charges for  $AdS_5 \times S^5$ and show that
its Penrose limit coincides with the charge previously found. We
also check explicitly that the semiclassical value for that $AdS$
charge for a rotating string on $S^5$ (the dual of a BMN state)
coincides with that of the pp-wave charge when applied to the same
BMN state. We conclude with some discussion and open problems.

\section{Nonlocal charges in coset models}

In this section we review the construction of non-local charges
for the Green-Schwarz superstring on $AdS_5 \times S^5$. We recall
the results found in \cite{Bena:2003wd} that will be useful in the
following and then we show that the Green-Schwarz superstring on
pp-waves possesses charges of the same structure.

\subsection{Nonlocal charges in $AdS_5 \times S^5$}

The Green-Schwarz superstring on $AdS_5 \times S^5$ can be
considered as a non-linear sigma model where the field $g(x)$
takes values in the coset superspace \cite{metsey,stanford,suny}:

\begin{equation}
\frac{G_{AdS}}{H_{AdS}}=\frac{PSU(2,2|4)}{SO(4,1)\times SO(5)},\
\label{supco}
\end{equation}
whose bosonic part is

\begin{equation}
\frac{SO(4,2)}{SO(4,1)} \times \frac{SO(6)}{SO(5)}=AdS_5 \times
S^5.
\end{equation}

The bosonic generators of $G_{AdS}$ are the translations $P^{a}$
and rotations $J^{ab}$, with $a,b=0,...,4$, (generators of
$SO(4,2)$) and translations $P^{a'}$ and rotations $J^{a'b'}$,
with $a',b'=0,...,4$, (generators of $SO(6)$). The fermionic
generators are 32 spinors $Q^{\alpha \alpha',I}$ with
$\alpha,\alpha'=1,...,4$ and $I=1, 2$. $H_{AdS}$ is the stability
subgroup of $G_{AdS}$, generated by the rotations $J^{ab}$ and
$J^{a'b'}$.

Then, it follows that the Lie algebra of $PSU(2,2|4)$ can be
decomposed in the following way:

\begin{equation} {\cal G}_{AdS} = {\cal H}_{AdS} + {\cal P} + {\cal Q}_1 + {\cal
Q}_2,
\end{equation}
with ${\cal H}_{AdS}$ the Lie algebra of $H_{AdS}$, ${\cal P}$ the
algebra of the translations, and  ${\cal Q}_1$ and  ${\cal Q}_2$
two copies of the ( 4, 4) of ${\cal H}_{AdS}$.

We focus on the current

\begin{equation}
\label{decomp}
 J = - g^{-1} \partial g = H + P + Q_1 + Q_2.
\end{equation}
Using the relation $dJ=J \wedge J$ and the ${\cal Z}_4$ grading
respected by the algebra \footnote{Under such grading ${\cal H}$,
${\cal P}$, ${\cal Q}_1$ and ${\cal Q}_2$ have charge 0, 2, 1 and
3 respectively.}, we can find equations for $dH$, $dP$, $dQ_1$ and
$dQ_2$. In terms of the lower-case currents, defined as
$x=gXg^{-1}$ they read: \footnote{ Capital letters, X, denote
currents that are conjugated by right multiplication, and they in
general correspond to some decomposition under representations of
the Lie Algebra, lower case letters, x, correspond to currents
conjugated by left multiplication and in general do not have a
simple decomposition under representations.}

\begin{eqnarray}
d h &=& -h \wedge h + p \wedge p - h \wedge p - p \wedge h - h
\wedge q - q \wedge h + {\textstyle
\frac{1}{2}}(q \wedge q - q' \wedge q')\ ,\nonumber\\
d p &=& - 2p \wedge p - p \wedge q - q \wedge p + {\textstyle
\frac{1}{2}}(q \wedge q + q'
\wedge q')\ ,\\
d q &=& -2q \wedge q\ ,\nonumber\\ d q' &=& -2p \wedge q' -2q'
\wedge p -q \wedge q' - q' \wedge q \ ,\nonumber \label{curls}
\end{eqnarray}
where we have defined $q=q_1+q_2$ and $q'=q_1-q_2$. This can be
supplemented with the equations of motion \cite{Metsaev:2001bj}

\begin{eqnarray} d {*p} &=&  p \wedge *q + *q \wedge p + {\textstyle
\frac{1}{2}}(q \wedge q' + q' \wedge q) \ ,\nonumber\\ 0 &=& p
\wedge (*q - {q'}) +  (*q - {q'}) \wedge p\ ,\\ 0 &=& p \wedge (q
- *{q'}) +  (q - *{q'}) \wedge p\ . \label{eom} \nonumber
\end{eqnarray}

Next, we define

\begin{equation}
a = \alpha p + \beta {*p} + \gamma q + \delta q',
\end{equation}
then, by requiring $a$ to be a flat connection, {\it i.e.} $da+a
\wedge a=0$, we find two one-parameter families of solutions,
given by

\begin{eqnarray}
\label{coef}
\alpha &=& -2 \sinh^2 \lambda\ ,\nonumber\\
\beta &=& \mp 2 \sinh\lambda \cosh\lambda\ ,\\
\gamma &=& 1 \pm \cosh \lambda\ , \nonumber\\
\delta &=& \sinh \lambda \ \nonumber .
\end{eqnarray}

Given a flat connection, the following equation

\begin{equation}
dU=-aU,
\end{equation}
is integrable, on a simply connected space, and with initial
condition $U(x_0,x_0)=1$, then

\begin{equation}
\label{wilson} U(x,x_0) = {\cal P} \exp \biggl(-\int_C a \biggr)\
\end{equation}
With $C$ a path from $x_0$ to $x$ and ${\cal P}$ the path ordering
in the Lie algebra. This allows the construction of an infinite
number of conserved charges, given by

\begin{equation}
\label{currents} Q^{a_\pm}(t) = U^{a_\pm}(\infty,t; -\infty, t).\
\end{equation}

This charges can be shown to be conserved, for an appropriate
falloff of the fields at infinity.

As we are interested in closed string theory, the world-sheet
respects periodic boundary conditions. Considering the product of
four Wilson lines that form a closed path in the strip (and don't
enclose any singularity) we have

\begin{equation}
U^{a_\pm}(0,t_0;0,t)U^{a_\pm}(0,t;1,t)U^{a_\pm}(1,t;1,t_0)=U^{a_\pm}(0,t_0;1,t_0),
\end{equation}
from this equation it is easy to see that the following quantity

\begin{equation}
\label{invcur} Q^{a_\pm}(t) = <U^{a_\pm}(0,t;1,t)>,
\end{equation}
is conserved. Here the spatial coordinate $\sigma$ is restricted
to the $[0,1]$ interval and we have assumed periodic boundary
conditions. With $< {\cal O} >$ we denote some
invariant cyclic operator. For the case of $AdS_5 \times S_5$,
whose algebra of isometries is a semi-simple algebra, the operator
$<>$ can simply be the trace.

A convenient way to see the infinite set of charges is by Taylor
expanding in the parameter $\lambda$, for instance

\begin{equation}
Q^{a_-}(t) = 1 + \sum_{n=1} \lambda^n Q_{n}.
\end{equation}

Writing $a=\lambda a^{(1)} + \lambda^2a^{(2)} + ...$, for the
first order charges we have

\begin{eqnarray}
Q_1=\int_0^1 {d \sigma a_1^{(1)}(\sigma)},\\
Q_2=\int_0^1 {d \sigma a_1^{(2)}(\sigma)}+ \int_0^1{d \sigma
\int_0^{\sigma}{d\sigma' a_1^{(1)}(\sigma) a_1^{(1)}(\sigma')}},
\end{eqnarray}
and so on.

\subsection{Nonlocal charges on the PP-Wave}

In this subsection we will argue that these non-local charges
exist also for the case of the pp-wave and have the same form
(\ref{invcur}).

Let us consider the Green-Schwarz superstring on a pp-wave RR
background

\begin{eqnarray}
\label{lin1}
 ds^2=2dx^+dx^- - m^2 x_I^2 dx^{+2} + dx_I^2\\
F^{-i_1\ldots i_4} = 2m \epsilon^{i_1\ldots i_4}\,, \hspace{0.3in}
F^{-i'_1 \ldots i'_4} = 2m \epsilon^{i'_1\ldots i'_4}\,
\end{eqnarray}
$I=1,\ldots 8$, describes the eight flat directions of the
pp-wave, $i,j=1,\ldots 4$ and $i',j'=5,\ldots 8$.. Here $m$ is a
dimensionful parameter, that from now on we will take equal to
one, and $x^+$ is taken to be the light-cone evolution parameter.

The Green-Schwarz superstring on a pp-wave background can be
regarded as a non linear sigma model on the coset superspace
\cite{Metsaev:2001bj}

\begin{equation}
\frac{G}{H}
\end{equation}

The transformation group $G$ is spanned by the following
generators: The even (bosonic) part of the superalgebra includes
ten translation generators $P^\mu$, $SO(4)$ rotation generators
$J^{ij}$, $i,j=1,\ldots, 4$, $SO'(4)$ rotation generators
$J^{i'j'}$, $i',j'=5,\ldots, 8$ and eight rotation generators in
the $(x^-,x^I)$ plane $J^{+I}$. The odd (fermionic) part of the
superalgebra consists of the complex 16-component spinor
$Q_\alpha$, $\alpha=1,\ldots 16$. The stability group $H$, is
generated by $J^{ij}$, $J^{i'j'}$ and $J^{+I}$. The algebra
between the relevant generators is given in the appendix.

Then, it follows that the Lie algebra ${\cal G}$ can be decomposed
as follows

\begin{equation}
{\cal G} = {\cal H} + {\cal P} + {\cal Q}_1 + {\cal Q}_2\ .
\end{equation}

Hence the current $J$ can be decomposed as in (\ref{decomp}). The
algebra in this case also respects a ${\cal Z}_4$ grading that
together with the condition $dJ=J \wedge J$ leads to equations of
the exact form of (\ref{curls}).

For simplicity, from now on we will deal explicitly with the
bosonic fields, however the discussion can be carried out for the
fermionic fields as well. The equation of motion reads
\cite{Metsaev:2001bj}:

\begin{equation}
d*p=0,
\end{equation}
which is the same as (\ref{eom}). We can then construct the same
set of nonlocal conserved charges (\ref{invcur}), from the
connection

\begin{equation}
a = \alpha p + \beta {*p},
\end{equation}
with $\alpha$ and $\beta$ given by (\ref{coef}). In terms of
Cartan 1-forms
\begin{eqnarray}
p=L^{\mu}p_{\mu},\hspace{0.3in} p_{\mu} =g P_{\mu} g^{-1},
\end{eqnarray}
with $L^{\mu}$ 1-forms in the two dimensional world-sheet and
$P_{\mu}$ generators of $G$. We know that string theory on this
background is exactly solvable in the light cone gauge, so it is
interesting to ask what is the form of these charges when we fix
such gauge.

\section{Explicit form of the charges on the PP-wave}

In the light cone gauge, the Cartan 1--forms become

\begin{equation}
\label{lcone}
 L^+=dx^+ , \hspace{0.3in} L^I=dx^I,\hspace{0.3in}
L^-=dx^- -\frac{1}{2}x_I^2 dx^+ , \end{equation} with $x^\mu$ two
dimensional fields on the world-sheet, depending on general on the
world-sheet coordinates $(\tau,\sigma)$ with metric $g_{ab}$.
Further we fix

\begin{equation}
\sqrt{g}g^{ab}=\eta^{ab},\hspace{0.3in}x^+(\tau,\sigma)=\tau,
\hspace{0.3in}p^+=1.
\end{equation}

With this, the flat connection takes the form

\begin{equation}
a=\alpha p + \beta (*p)= \pm 2 \lambda (*p) - 2 \lambda^2 p \pm
\frac{4}{3}\lambda^3 + ...,
\end{equation}
with

\begin{equation}
\label{pbpf} p = d \tau p_+ + dx^I p_I + (dx^- -\frac{1}{2} x_I^2
d \tau) p_{-}.
\end{equation}

It is important to notice that $p$ is written in terms of
lower-case generators, of the form $p^{\mu}=g P^{\mu} g^{-1}$, and
hence dependent on the world-sheet coordinates via $g$. Choosing a
coset representative of the form $g(x^{\mu})=e^{\tau P^-}e^{x^I
P^I}$ we can express the lower-case generators in terms of
upper-case generators as follows

\begin{eqnarray}
p^I=\cos{\tau} P^I - \sin{\tau} J^{+I},\\
p^{-}=P^{-}-\frac{x^I x^I}{2}P^+ + x^I \cos{\tau} J^{+I}+ x^I \sin{\tau} P^{I},\\
p^+=P^+.
\end{eqnarray}

Writing these charges in terms of upper-case generators, we see
their explicit dependence on the world-sheet coordinates, then by
Taylor expanding  in the parameter $\lambda$ we can construct the
charges order by order.

\subsection{First order charge}

For the first order in $\lambda$ we obtain the charge

\begin{eqnarray}
Q_1=\left< \int_{0}^{1}{d \sigma (P^{-} + A(\sigma)P^{+}
+B^I(\sigma)P^I +C^I(\sigma)J^{+I} )} \right>,
\end{eqnarray}
with

\begin{eqnarray}
 A(\sigma)=\partial_{\tau} x^{-} - x^I x^I\\
B^I(\sigma)=\sin{\tau} x^I + \cos{\tau}\partial_{\tau} x^{I}\\
C^I(\sigma)=\cos{\tau} x^I - \sin{\tau}\partial_{\tau} x^{I}
\end{eqnarray}
where we are not showing the $\sigma$ dependence of the
world-sheet fields. By using (\ref{xm}) we can write $A(\sigma)$
purely in terms of the fields $x^I$
\begin{equation}
A(\sigma)=-\frac{1}{2}(\partial_\tau x^I \partial_\tau
x^I+\partial_\sigma x^I \partial_\sigma x^I + x^I x^I).
\end{equation}

By using the equations of motion, we can see that the coefficient
of every generator is a classically conserved charge. So we have
four conserved quantities

\begin{equation}
Q_1=1,\hspace{0.3in}Q_A=\int_{0}^{1}{d \sigma
A(\sigma)},\hspace{0.3in}Q_{B^I}=\int_{0}^{1}{d \sigma
B^I(\sigma)},\hspace{0.3in}Q_{C^I}=\int_{0}^{1}{d \sigma
C^I(\sigma)}.
\end{equation}

Since every coefficient is conserved, this implies the
conservation of $Q_1$, for every invariant $<>$ we choose.

It is interesting to notice that if we plug the mode expansion of
the fields $x^I$ (see appendix) we find the following expressions
for the charges

\begin{eqnarray}
\label{q1comp}
Q_{A}=\frac{1}{2}\left( p_0^Ip_0^I+ x_0^Ix_0^I \right)+\sum_{n \ne 0}\left( \alpha_{n}^{I 1}\alpha_{-n}^{I 1} +\alpha_{n}^{I 2}\alpha_{-n}^{I 2} \right),\\
Q_{B^I}=p_0^I,\hspace{0.3in}Q_{C^I}=x_0^I.
\end{eqnarray}

The classical Poisson-Dirac brackets among these charges are given
by

\begin{equation}
[Q_{B^I},Q_{C^J}]=\delta^{IJ}
Q_T,\hspace{0.3in}[Q_{B^I},Q_{A}]=Q_{C^I},\hspace{0.3in}[Q_{A},Q_{C^I}]=Q_{B^I},
\end{equation}
whereas, of course, $Q_T$ has zero Poisson-Dirac bracket with
every operator. Note that this is the same algebra followed by the
bosonic generators of the pp-wave algebra.

>From (\ref{q1comp}) we see that these three charges represent the
constants of motion $p_0^I$, $x_0^I$ and the Hamiltonian, that
are, of course, the quantities associated with the symmetries of
the pp-wave.

\subsection{Second order charge}

For the second order charge we have \footnote{$Q_2$ has in general
a single integral term, proportional to $\int{p_1}$, however, this
is 0 in this case, since $p_1$ is a total derivative in the
spatial coordinate.}

{\footnotesize
\begin{eqnarray}
Q_2=\left< \int_{0}^{1}{d \sigma \int_{0}^{\sigma}{d \sigma'(P^{-}
+ A(\sigma)P^{+} +B^I(\sigma)P^I +C^I(\sigma)J^{+I} )(P^{-} +
A(\sigma')P^{+} +B^J(\sigma')P^J +C^J(\sigma')J^{+J} )}}
\right>\nonumber
\end{eqnarray}
}

In order to study $Q_2$ let us introduce the following notation:

\begin{equation}
Q_{{\cal AB}}=\int_{0}^{1}{d \sigma \int_{0}^{\sigma}{d
\sigma'{\cal A}(\sigma) {\cal B}(\sigma')}},
\end{equation}
where ${\cal A , B}$ can take the values $1, A, B^I$ or $C^I$.
Then we find the following conserved quantities

\begin{eqnarray}
Q_{11},\hspace{0.25in}Q_{AA},\hspace{0.25in}Q_{BB},\hspace{0.25in}Q_{CC}\nonumber\\
Q_{\{1,A\}}=Q_{1A}+Q_{A1},\hspace{0.25in}Q_{\{1,B\}}=Q_{1B}+Q_{B1},\hspace{0.25in}Q_{\{1,C\}}=Q_{1C}+Q_{C1},\nonumber\\
Q_{\{A,B\}}=Q_{BA}+Q_{AB},\hspace{0.25in}Q_{\{A,C\}}=Q_{AC}+Q_{CA},\hspace{0.25in}Q_{\{B,C\}}=Q_{BC}+Q_{CB},\nonumber\\
Q_a=\frac{1}{2}Q_{[C^I,B^I]} <P^+> +Q_{[B^I,1]}
<J^{+I}>+Q_{[1,C^I]} <P^I>.\nonumber
\end{eqnarray}
where we have introduced the notation $Q_{[{\cal A},{\cal B}]}= Q
_{{\cal AB}}- Q _{{\cal BA}}$. The charges in the first three
lines, can be written as product of the charges appearing in
$Q_1$, in fact

\begin{eqnarray}
Q_{{\cal \{ A,B \} }}=\int_{0}^{1}{d \sigma \int_{0}^{\sigma}{d \sigma'{\cal A}(\sigma) {\cal B}(\sigma')}}+\int_{0}^{1}{d \sigma \int_{0}^{\sigma}{d \sigma'{\cal B}(\sigma) {\cal A}(\sigma')}}= \nonumber\\
=\int_{0}^{1}{d \sigma \int_{0}^{1}{d \sigma'{\cal A}(\sigma)
{\cal B}(\sigma')}}=Q_{{\cal A }} Q_{{\cal B }},
\end{eqnarray}
where we have used that ${\cal A}$ and ${\cal B}$ commute. As for
the last charge, note that since the invariant operator in cyclic,
then $<T^A T^B>-<T^BT^A>=<\left[ T^A,T^B \right]>=0$ and it
becomes trivial.

In order to get some non-trivial conserved charge we should go to
higher order.

\subsection{Third and higher order charges}

In this subsection we try to determine the general structure of
the higher order charges and give an algorithm to write them
explicitly.

As seen in the previous section, the charge of order $N$ will have
a local  contribution (only one integral) plus a bi-local (two
integrals), etc,  up to a N-local contribution, given
schematically by sum of terms of the form

\begin{equation}
\label{qn} Q_{A_1...A_N}=\int_0^1 d \sigma_1 {\cal A}_1
\int_0^{\sigma_1} d \sigma_2 {\cal A}_2 ...\int_0^{\sigma_{N-1}} d
\sigma_N {\cal A}_N,
\end{equation}
times the corresponding product of generators, plus all the
permutations.

In order to study these combinations, let us notice that we can
write an arbitrary permutation in the following form
\begin{eqnarray}
\label{permqn}
Q_{A_{I_1}...A_{I_N}}&=&\int_0^1 d \sigma_1 {\cal A}_1 \int_{\sigma_1^i}^{\sigma_1^{i+1}} d \sigma_2 {\cal A}_2 ...\int_{\sigma_{N-1}^{j}}^{\sigma_{N-1}^{j+1}} d \sigma_N {\cal A}_N\\
Q_{A_{I_1}...A_{I_p}BA_{I_{p+1}},...A_{I_N}}&=&\int_0^1 d \sigma_1
{\cal A}_1 \int_{\sigma_1^i}^{\sigma_1^{i+1}} d \sigma_2 {\cal
A}_2 ...\int_{\sigma_{N-1}^{j}}^{\sigma_{N-1}^{j+1}} d \sigma_N
{\cal A}_N  \int_{\sigma_{N}^{N-p}}^{\sigma_{N}^{N-p+1}} d
\sigma_{N+1} {\cal B}, \nonumber
\end{eqnarray}
with $\sigma_m^i$ taking the values $0, \sigma_1, ..., \sigma_m,
1$ in crescent order. In other words, we express the permutations
by interchanging the intervals of integration instead of the order
of the ${\cal A}_i$.

We can give a precise recursive relation giving the integral of
any permutation if we complement (\ref{permqn}) with the following
relation

\begin{equation}
\int_0^1 d \sigma {\cal B}(\sigma) \int_{0}^{\sigma} d \sigma_1
{\cal A}(\sigma_1) = \int_0^1 d \sigma {\cal A}(\sigma)
\int_{\sigma}^1 d \sigma_1 {\cal B}(\sigma_1).
\end{equation}

For such a N-local integral, we have $N!$ permutations, however,
not all of them are independent of integrals appearing at lower
order. For instance, with the recursive relation given here, it is
easy to see that the completely symmetric sum of all the
permutations is just the product of the local integrals

\begin{equation}
\label{perm} Q_{A_1...A_N} + \hbox{permutations} =Q_{A_1} ...
Q_{A_N}.
\end{equation}

More generally, one can prove that

\begin{eqnarray}
\label{relq}
Q_{A_1...A_{N}B}+Q_{A_1...A_{N_1}BA_N}+...+Q_{BA_1...A_N}=Q_BQ_{A_1...A_N},
\end{eqnarray}
from where (\ref{perm}) as well as other relations can be shown.
By using (\ref{relq}) together with the commutation relation among
the generators one can write an arbitrary order charge in a way in
which lower order contributions are explicit. For instance, for
the third order charge we have \footnote{There will be also a
local term, that is proportional to $Q_1$, and a bi-local term,
whose contribution vanish for the case under consideration}

\begin{eqnarray}
\label{nlocal} Q_3=Q_{ABC} \left< P^A P^B P^C \right> =
\frac{1}{6} \left( Q_{ABC} \left< P^A P^B P^C \right> +
\hbox{permutations} \right)=\nonumber\\=\frac{1}{12}Q_AQ_BQ_C
\left< P^A \{P^B,P^C \}\right> + \frac{1}{12}Q_AQ_{[B,C]} \left<
P^A [P^B,P^C]\right>+\\+\frac{1}{6}Q_{A[B,C]} \left< P^A
[P^B,P^C]\right> =\nonumber\\= \frac{1}{12}Q_AQ_BQ_C \left< P^A
\{P^B,P^C \}\right> + \frac{1}{4}Q_AQ_{[B,C]} \left< P^A
[P^B,P^C]\right>.\nonumber
\end{eqnarray}

As happened for $Q_2$, the first contribution of the right-hand
side of (\ref{nlocal}) is conserved, independently of the choice
of $<>$, since it is the product of conserved charges. Let us
focus in the nontrivial piece

\begin{eqnarray}
Q^{NT}_3=Q_{A}Q_{[B,C]}\left< P^A, [P^B,P^C]\right>=Q_{A}Q_{[B,C]}
f^{BC}_D \left< P^A P^D \right>.
\end{eqnarray}

At this point we need to give an expression for $\left< P^A P^D
\right>$, that we will call $\Omega^{AD}$. In our case $P^A$ can
take the values $P^I, J^{+I}, P^-$ and $P^+$. If we take
$\Omega^{AB}=Tr(P^A P^B)$ , then we obtain

$$\Omega^{AB}=\begin{pmatrix}0&0&0&0\cr
0&0&0&0\cr 0&0&-2&0\cr 0&0&0&0 \end{pmatrix}$$
and as can be
easily seen, we obtain trivial conserved charges. The fact that
$\Omega^{AB}$ is degenerate is due to the fact the pp-wave algebra
is non semi-simple. Fortunately, the most general $\Omega^{AB}$
with the required properties has been given for the algebra under
consideration \cite{inv}\cite{nw} \footnote{For the product of
$P^IP^J$, or $J^{+I}J^{+J}$, we take the invariant to be
proportional to $\delta^{IJ}$.}

$$\Omega^{AB}=\begin{pmatrix}k&0&0&0\cr
0&k&0&0\cr 0&0&b'&k\cr 0&0&k&0\end{pmatrix}$$

With this choice one can see that $Q^{NT}_3$ is non trivial and
conserved:

\begin{eqnarray}
Q^{NT}_3= Q_{[C^I,B^I]}Q_1+Q_{[B^I,1]}Q_{C^I}+Q_{[1,C^I]}Q_{B^I}=\nonumber \\
=-2\int_{0}^{1}{d \sigma \int_{0}^{1}{d \sigma' \sigma x^I(\sigma) \partial_{\tau} x^I(\sigma')}}+2\int_{0}^{1}{d \sigma \int_{0}^{1}{d \sigma' \sigma \partial_{\tau} x^I(\sigma) x^I(\sigma')}} +\\
+\left( \int_{0}^{1}{d \sigma \int_{0}^{\sigma}{d \sigma'
x^I(\sigma) \partial_{\tau} x^I(\sigma')}}-\int_{0}^{1}{d \sigma
\int_{0}^{\sigma}{d \sigma' \partial_{\tau} x^I(\sigma)
x^I(\sigma')}}\right).\nonumber
\end{eqnarray}

Written in terms of the mode expansion

\begin{equation}
Q^{NT}_3=\sum_{n \ne 0}\frac{2}{w_n k_n}\left( \alpha_{n}^{I
1}\alpha_{-n}^{I 1} -\alpha_{n}^{I 2}\alpha_{-n}^{I 2} \right).
\end{equation}

In order to evaluate higher order charges, one should give an
expression for higher order invariants. In general they will have
contribution from lower order invariants, plus some independent
piece. We stress that in general, for a given order charge, there
will be terms conserved by themselves, for instance, at fourth
order we find

\begin{eqnarray}
Q^{I}_4=\int_0^1 d \sigma x^I(\sigma) \int_0^{\sigma} d \sigma'
x^I(\sigma') + \int_0^1 d \sigma \partial_{\tau} x^I(\sigma)
\int_0^{\sigma} d \sigma' \partial_\tau x^I(\sigma'),
\end{eqnarray}
plus some other complicated contributions. Note that even if the
complete charges of a given order is a Casimir of the group, there
will be components that are conserved by themselves and need not
to be a Casimir. Plugging the oscillator expressions for the
fields $x^I$ we obtain

\begin{eqnarray}
Q^{I}_4= \frac{1}{2}(x_0^Ix_0^I+p_0^Ip_0^I) - i \sum_m
\frac{k_n}{w_n^2}(\alpha_n^1\alpha_{-n}^1
-\alpha_n^2\alpha_{-n}^2).
\end{eqnarray}

From the oscillator expressions for $Q_3^{NT}$ and $Q^{I}_4$ it is
evident that they will have vanishing classical Poisson Dirac
brackets between them, so we see that they will not generate new
charges. This is different to the situation of sigma models with
boundary conditions at infinity, where all the infinite set of
classically conserved non-local charges is generated by the first
non-local charges.

As we will see in the next section, there is another procedure to
recover such charges, that simply consist in taking the Penrose
limit of the corresponding charges on  $AdS_5 \times S^5$. In this
case the algebra is semi-simple, and so we can take the invariant
operator as the trace.

\section{$AdS_5 \times S^5$ charges and their Penrose limit}

In the previous section we have given the explicit form for the
infinite set of non-local classically conserved charges for the
pp-wave by studying directly string theory on such background. It
is expected that such charges are the Penrose limit of the charges
for the $AdS_5 \times S^5$. In this section we will prove that
this is the case for the first non-local charge.

\subsection{Explicit form of the Charges on $AdS_5 \times S_5$}

As before, the first order charge can be written as (we are
forgetting about the trace)

\begin{equation}
Q^{1}_{AdS}=\int(L_a p^a + L_{a'} p^{a'}),
\end{equation}
with the Cartan 1-forms given by \cite{hks}

\begin{eqnarray}
L^a_{\rm AdS}=dy^a+\left(\displaystyle\frac{\sinh
y}{y}-1\right)dy^b{\bf \Upsilon}_b^{~~a} &,& L^{a'}_{\rm
AdS}=dy^{a'}+\left(\displaystyle\frac{\sin
y'}{y'}-1\right)dy^{b'}{\bf \Upsilon}_{b'}^{~~a'} \label{AdSe},
\end{eqnarray}
with

\begin{eqnarray}
y=\sqrt{y^2}=\sqrt{y^ay_a},\hspace{0.3in}
y'=\sqrt{{y'}^2}=\sqrt{{y}^{a'}y_{a'}},\nonumber\\
{\bf
\Upsilon}_{a}^b=\delta_a^{~b}-\displaystyle\frac{y_ay^b}{y^2}~~,
\hspace{0.3in} {\bf
\Upsilon}_{a'}^{b'}~=~\delta_{a'}^{~b'}-\displaystyle\frac{y_{a'}y^{b'}}
{{y'}^2}~~\label{projectionsb}~~~.
\end{eqnarray}

In this coordinates the bosonic Lagrangian (equivalently the
metric) reads:

\begin{eqnarray}
{\cal L}= (dy)^2+(\sinh{y})^2 d \Omega_4^2 + (dy')^2+(\sin{y'})^2
d'\Omega_{4'}^2
\end{eqnarray}
where the 4-sphere metrics are given by

\begin{eqnarray}
d \Omega_4^2 = \frac{(dy^a)(dy^a)-(dy)^2}{y^2},
\end{eqnarray}
and the same for $d'\Omega_{4'}^2$. As before, in order to show
the explicit dependence on the world-sheet coordinates it is
convenient to express lower-case generators in terms of upper-case
generators, for the present case we obtain

\begin{eqnarray}
p^{a}=\cosh{y}P^{a}+\left(\frac{1-\cosh{y}}{y^2}\right) y^{a}y^{b}P^{b}-\frac{\sinh{y}}{y}y^{b}J^{ab},\\
p^{a'}=\cos{y'}P^{a'}+\left(\frac{1-\cos{y'}}{y'^2}\right)
y^{a'}y^{b'}P^{b'}+\frac{\sin{y'}}{y'}y^{b'}J^{a'b'}.
\end{eqnarray}

Now in terms of upper-case generators

\begin{eqnarray}
Q^{1}_{AdS}=\int(C_a P^a +C_{ab} J^{ab}+ C_a' P^{a'}+C_{a'b'}
J^{a'b'}),
\end{eqnarray}
with

\begin{eqnarray}
C_a= \cosh{y} L_a + \left(\frac{1-\cosh{y}}{y^2} \right) y^b dy^b y^a,\hspace{0.2in} C_{ab}=-\frac{\sinh{y}}{y}\left( L^a y^b - L^b y^a \right),\\
C_{a'}= \cos{y'} L_{a'} + \left(\frac{1-\cos{y'}}{y'^2} \right)
y^{b'} dy^{b'} y^{a'},\hspace{0.2in}
C_{a'b'}=\frac{\sin{y'}}{y'}\left( L^{a'} y^{b'} - L^{b'} y^{a'}.
\right)\end{eqnarray}

So the charges $\int_0^1{d\sigma C_{a,0}}$, etc, are conserved
quantities, and they represent the isometries of theory. As it is
well know, by performing the Penrose limit, the isometries of
$AdS_5 \times S_5$ map into the isometries of the pp-waves, so the
Penrose limit of this first order charges are the first order
charges found previously.

In order to find higher order charges we should worry about the
invariant $<>$. The algebra of $AdS_5 \times S_5$ is semi-simple,
so we can simply take the trace of product of upper-case
operators. Again the second order charge will be trivial and we
should go to the third order.

Let us write the upper-case generators as $T^a$, then
$[T^a,T^b]=f^c_{ab}T^c$. Next, let us choose a representation of
the algebra in which $Tr(T^a T^b) \propto \delta_{ab}$ then the
non trivial third order charge becomes

\begin{equation}
Q^3_{AdS}=f^a_{bc}\int_0^{1}{C_a}\left( \int_0^1{d\sigma
C_b(\sigma) \int_0^\sigma C_c(\sigma') }\right),
\end{equation}
where now $C_a$ is the coefficient of $T^a$ in the first order
charge, etc.

\subsection{Penrose limit}

The Penrose limit of the Cartan 1-forms were done in \cite{hks},
where we refer the reader for the details, here we repeat
basically their analysis. \footnote{Our conventions interchange +
and - with respect to the conventions used in \cite{hks}.} We
define $y^{\mp}=y_{\pm}=(y^9 \pm y^0)/ \sqrt{2}$ and then perform
the rescaling

\begin{eqnarray}
y^-\to \Omega^2 y^-,~~y^+\to y^+,~~y^{\hat{i}}\to \Omega
y^{\hat{i}}.
\end{eqnarray}

The Cartan 1-forms should also be rescaled as $L^- \rightarrow
\Omega^2 L^-$, $L^+ \rightarrow L^+$ and  $L^{\hat{i}} \rightarrow
\Omega L^{\hat{i}}$.

Finally by performing the following change of coordinates

\begin{eqnarray}
x^{\hat{i}}=\frac{\sin{y^+}}{y^+}y^{\hat{i}},\\
x^+=y^+,\\
x^-=y^- +
\frac{y^{\hat{i}}y^{\hat{i}}}{2y^+}\left(1-\frac{\sin{2y^+}}{2y^+}\right),
\end{eqnarray}
we obtain the Cartan 1 forms used in the previous section for the
pp-wave as $\Omega \rightarrow 0$ (see (\ref{lcone})).

\begin{eqnarray}
L^-=dx^- -\frac{1}{2}x^{\hat{i}}x^{\hat{i}}
dx^+,\hspace{0.3in}L^+=dx^+,\hspace{0.3in}L^{\hat{i}}=dx^{\hat{i}}.
\end{eqnarray}

In order to show that $Q^3_{AdS}$ maps into $Q_3^{NT}$ we need
also to show that the $AdS_5 \times S_5$ algebra goes to the
pp-wave algebra, with the correct structure constants $f^A_{BC}$.
This was done in \cite{hks2}, by performing the following
rescaling

\begin{eqnarray}
P^+ \rightarrow \frac{1}{\Omega^2}P^+,\hspace{0.3in}P^{\hat{i}}
\rightarrow
\frac{1}{\Omega}P^{\hat{i}},\hspace{0.3in}P^{\hat{i}}_*
\rightarrow \frac{1}{\Omega}P^{\hat{i}}_*,
\end{eqnarray}
where $P_*^a$ are the boost generator, and then taking the $\Omega
\rightarrow 0$ limit. Notice that these rescaling corresponds to
the rescaling on the coordinates.

So we see that  $Q^3_{AdS}$ maps into $Q_3^{NT}$. It is
interesting to notice that to construct the charges in  $AdS_5
\times S^5$, where we can take the trace of products of operators
as invariant form without loosing generality, since the algebra is
semi-simple, and then to take their Penrose limit, it is
equivalent to consider the charges on the pp-wave but using now
the non-degenerate invariant, as done in the previous section.

\subsection{An explicit check}

As an explicit check that the Penrose limit of the charges of
$AdS_5 \times S^5$ are the charges on the pp-wave we can consider
the following exercise.

Since we have the expression for $Q_3^{NT}$ in terms of the mode
expansion of the coordinates $x^I$, we know what is its value when
applied to BMN operators \footnote{For future convenience, we use
the notation used in \cite{bmn}, where $n>0$ for left movers and
$n<0$ for right movers.}:

\begin{eqnarray}
\label{q3val} Q_3^{NT} \alpha^{\dagger}_{n_1}
\alpha^{\dagger}_{n_2}... \alpha^{\dagger}_{n_L}|0> \sim
\left(\frac{1}{n_1}+\frac{1}{n_2}+...+\frac{1}{n_L}\right)\alpha^{\dagger}_{n_1}
\alpha^{\dagger}_{n_2}... \alpha^{\dagger}_{n_L}|0>,
\end{eqnarray}
where we have taken the classical limit, {\it i. e.} $\sqrt{1+n^2}
\approx 1$ \footnote{ More explicitly, reintroducing the
dimensionful parameters $\sqrt{m^2+\frac{n^2}{(\alpha' p^+)^2}}
\approx m$.}. On the other hand, on $AdS_5 \times S_5$, the dual of
the BMN operators are believed to be rotating strings on an
equator of $S^5$ with very large angular momentum. So it is
interesting to check the value of  $Q^3_{AdS}$ when we plug on it
the semiclassical solution corresponding to such rotating string.
This was done for the hamiltonian in \cite{gkp}(see also \cite{russo}).

In the following we will focus on the $S^5$ part, the
analysis for the $AdS_5$ is very much the same. First, let us
change coordinates to the one used by \cite{bmn}. In such coordinates,
the metric turns out to be

\begin{eqnarray}
ds^2=d\phi^2 \cos^2 \theta +d\theta^2 +\sin^2\theta d \Omega_3^2,
\end{eqnarray}
with $d \Omega_3^2$ the metric of a 3-sphere parametrized by
angles $\alpha$, $\beta$ and $\gamma$. Performing such change of
coordinates and setting $\alpha=\beta=\gamma=0$ for simplicity we
can rewrite the coefficients $C_a$ in terms of the new coordinates
(As we focus on the $S^5$ only we will suppress the primes)

\begin{eqnarray}
C_{J^{45}}=-C_{J^{54}}=\cos{\phi}d\theta +\sin{\theta}\cos{\theta}\sin{\phi} d \phi,\\
C_{P^4}=\sin{\phi}d\theta -\sin{\theta}\cos{\theta}\cos{\phi} d \phi,\\
C_{P^5}=-\cos^2\theta d \phi.
\end{eqnarray}

We will consider a string rotating in the equator defined by
$\theta=0$, that is, we will consider $\phi= J \tau$ and small
oscillations around $\theta=0$. The Lagrangian becomes

\begin{eqnarray}
{\cal L} = \cos^2 \theta (d \phi)^2 +(d \theta)^2 \approx
J^2(1+\theta^2) +(d \theta)^2
\end{eqnarray}

By using the approximation of small perturbation and the explicit
form of $\phi$ we obtain

\begin{eqnarray}
\label{adscharges}
j^A \equiv C_{J^{45}}=-C_{J^{54}}=\cos{J \tau}\, d\theta +J \sin{J \tau} \,\theta d \tau,\nonumber\\
j^B \equiv C_{P^4}=\sin{J \tau}\, d\theta -J \cos{J \tau} \,\theta d \tau, \\
j^C \equiv C_{P^5}=- J d \tau.\nonumber
\end{eqnarray}

By using the equation of motion

\begin{equation}
(\partial_\sigma^2-\partial_\tau^2)\theta-J^2\theta=0,
\end{equation}
it is easy to see that the quantities $Q^A=\int_0^1{j^A_0}$, etc,
are conserved. Let us suppose a perturbation of the form \footnote{The presence
of 0-modes will not change the final result.}

\begin{eqnarray}
\theta(\tau,\sigma)= \sum_{i=1}^{K} \frac{A_i}{\omega_{n_i}}\left(
e^{i(\omega_{n_i} \tau + 2 \pi n_i \sigma)}+e^{-i(\omega_{n_i}
\tau + 2 \pi n_i \sigma)} \right)
\end{eqnarray}
with $\omega_n = \sqrt{J^2+4 \pi n^2}$. Then we find

\begin{eqnarray}
Q_{AdS}^3=Q^{A}Q^{[B,C]} +Q^{C}Q^{[A,B]}+ Q^{B}Q^{[C,A]} = \nonumber\\= J(
\int_0^1{\partial_\tau \theta} \int_0^\sigma{\theta} -
\int_0^1{\theta} \int_0^\sigma{\partial_\tau \theta} )=
\sum_{i=1}^{K}\frac{J^2 A_i^2}{n_i\sqrt{J^2+4 \pi n_i^2}},
\end{eqnarray}
that coincides with (\ref{q3val}) for large J. In fact, one can
notice that the form of (\ref{adscharges}) coincides with that of
the coefficients $B^I$, $C^I$ and 1, of the pp-wave, from which
$Q_3^{NT}$ is built.

Of course this same method can be used to compute this charges
for other string states on $AdS_5 \times S^5$ , as done in
\cite{gkp}.

\section{Conclusions}

Recently in \cite{Bena:2003wd} it was found that the Green-Schwarz
superstring on $AdS_5 \times S^5$ possesses an infinite set of non-local classically conserved charges. This would suggest that in
some non-trivial cases the world-sheet theory may be exactly
solvable. To understand the role of these charges, their gauge
dual, etc, seems a very complicated task. As a warm up exersice we
propose the study of such charges on the pp-wave limit.

In this paper we show that the closed superstring on pp-waves
possesses an infinite tower of non-local classically conserved
charges. We then show that they are the Penrose limit of the
charges present for the $AdS_5 \times S^5$ background.

In order to construct these charges in closed string theory, one
must impose periodic boundary conditions on the world-sheet fields
and then take an invariant of the group element appearing in the
charge. As a consequence it is not clear whether it is possible to
generate all the tower of non-local charges by repeated Poisson
Dirac brackets of the first non-local charges (or some finite number of them), as opposed to what
happens when one considers the uncompactified sigma model. Indeed,
from the first order charges explicitly obtained in this paper it
is not possible to obtain more non-local conserved quantities.

On the other hand, when one considers closed string theory on
pp-waves, which has a non semi-simple algebra, the naive
invariant, {\it i. e. } the trace, turns out to be degenerate, and
one should look for a non degenerate invariant in order to obtain
non-trivial charges. The non degenerate bilinear invariant for the
algebra under consideration was found in \cite{inv}\cite{nw} and we use it in order to
compute the first non trivial non-local conserved charge.
Remarkably, this charge coincides with the Penrose limit of the
first non trivial non-local conserved charge for  $AdS_5 \times
S^5$, whose algebra is semi-simple and we can use the trace as
non-degenerate invariant.

There are many possible further directions to pursue. One could
try to complete the analysis for the fermionic sector of the
theory, here the simplification obtained by fixing the light cone
gauge is more relevant. From the discussion done in this paper, it
seems that we cannot generate all the infinite tower of non-local
charges from the first non-local charges, at least not as in the
case of unbounded sigma model, it could be interesting to show that
this is the case in general when one considers closed string
theory, or to show how to generate the tower from a finite number
of charges.  In the case in which the tower cannot be generated one
should develop more effective methods that evaluanting the charges
order by order. The question about the gauge dual of such charges
is interesting. As the AdS/CFT correspondence is more presice in
the pp-wave limit (as string theory is exactly solvable in this
background) maybe simpler to think about the dual of the charges
in this limit.

Even though string theory is exactly solvable on pp-waves, the
non-local charges constructed in this paper could provide a clue
about the role played by these charges in the full $AdS_5 \times
S^5$ background.

\noindent{\bf Acknowledgments}

I would like to thank J. David, E. Gava and K. Narain for useful
comments and discussions at every stage of this work. I would like
to thank also F. Bigazzi, M. Cirafici, C. Maccaferri, L.
Mazzucato,  P. Meessen and J. Russo.

\appendix

\section{Notation}

The commutation relations between the bosonic generators of the
$AdS_5 \times S^5$ algebra are

\begin{eqnarray}
\left[ P^a,P^b \right]=J^{ab}\hspace{0.4in}\left[ P^{a'},P^{b'}\right]=-J^{a'b'},\\
\left[ P^a,J^{bc} \right]=\eta^{ab}P^c-\eta^{ac}P^b,\hspace{0.4in}\left[ P^{a'},J^{b'c'}\right]=\eta^{a'b'}P^{c'}-\eta^{a'c'}P^{b'},\\
\left[ J^{ab},J^{cd} \right]=\eta^{bc}J^{ad}+\hbox{3
terms}\hspace{0.4in}\left[ J^{a'b'},J^{c'd'}
\right]=\eta^{b'c'}J^{a'd'}+\hbox{3 terms},
\end{eqnarray}
with $a=0,...,4$, so(4,1) vector indices, in the tangent space of
$AdS_5$, $a'=0,...,4$ so(5) vector indices in the tangent space of
$S^5$, $\eta^{ab}=diag(-++++)$ and  $\eta^{a'b'}=diag(+++++)$.

The commutation relations between the bosonic generators of the
pp-wave algebra are

\begin{equation}
[P^-,P^I] = -J^{+I},
\end{equation}
\begin{equation}
[P^I,J^{+J}] = - \delta^{IJ} P^+ ,\hspace{0.3in} [P^-,J^{+I}]=P^I,
\end{equation}
with $I=1,...,8$. These generators admit the following field
representation

\begin{eqnarray}
P^+=1, \hspace{0.3in} P^I=-\int_0^1{d \sigma(\cos \tau \partial_\tau x^I + \sin \tau x^I)},\\
 J^{+I}=\int_0^1{d \sigma(\sin \tau \partial_\tau x^I - \cos \tau x^I)}.
\end{eqnarray}

The two dimensional fields $x^I(\tau,\sigma)$ satisfy the
following equations of motion and periodicity conditions:

\begin{eqnarray}
(- \partial^2_{\tau}+ \partial^2_{\sigma})x^I-x^I=0\\
x^I(\tau,0)=x^I(\tau,1), \hspace{0.3in} \partial_\sigma x^I(\tau,0)= \partial_\sigma x^I(\tau,1)
\end{eqnarray}

Such equations admit as solution

\begin{equation}
\label{xIsol} x^I(\sigma,\tau) = \cos \tau\, x_0^I + \sin \tau\,
p_0^I + {\rm i}\sum_{n \neq 0} \frac{1}{\omega_n} e^{-{\rm i}
\omega_n \tau}(e^{{\rm i}k_n \sigma} \alpha_n^{1 I}+e^{-{\rm i}k_n
\sigma} \alpha_n^{2 I}),
\end{equation}
with the frequencies defined by

\begin{eqnarray}
\omega_n = \sqrt{k_n^2+1}, \hspace{0.3in} n>0; \hspace{0.3in}\omega_n = -\sqrt{k_n^2+1}, \hspace{0.3in} n<0,\\
k_n = 2 \pi n, \hspace{0.3in} n = \pm 1, \pm 2, ...
\end{eqnarray}

The coordinate $x^-$ can be expressed in terms of $x^I$ by the
following constraints

\begin{eqnarray}
\partial_\sigma x^-=-\partial_\tau x^I \partial_\sigma x^I,\\
\partial_\tau x^-=\frac{1}{2}(-\partial_\tau x^I \partial_\tau x^I-\partial_\sigma x^I \partial_\sigma x^I + x^I x^I)\label{xm}.
\end{eqnarray}

\end{document}